# Automated Corrosion Detection Using Crowd Sourced Training for Deep Learning

W.T. Nash[1,*], C.J. Powell[1], T. Drummond[2] and N. Birbilis[3]

[1]Department of Materials Science and Engineering, Monash University, Clayton. VIC. 3800. Australia.
[2]Department of Electrical and Computer Systems Engineering, Monash University, Clayton. VIC. 3800. Australia.
[3]College of Engineering and Computer Science, The Australian National University, Acton. ACT. 2601. Australia.

*will.nash@monash.edu

**Abstract:**

The automated detection of corrosion from images (i.e., photographs) or video (i.e., drone footage) presents significant advantages in terms of corrosion monitoring. Such advantages include access to remote locations, mitigation of risk to inspectors, cost savings and monitoring speed. The automated detection of corrosion requires deep learning to approach human level artificial intelligence (A.I.). The training of a deep learning model requires intensive image labelling, and in order to generate a large database of labelled images, crowd sourced labelling via a dedicated website was sought. The website (corrosiondetector.com) permits any user to label images, with such labelling then contributing to the training of a cloud based A.I. model – with such a cloud-based model then capable of assessing any fresh (or uploaded) image for the presence of corrosion. In other words, the website includes both the crowd sourced training process, but also the end use of the evolving model. Herein, the results and findings from the website (corrosiondetector.com) over the period of approximately one month, are reported.

**Keywords:** corrosion, corrosion detection, monitoring, machine learning, artificial intelligence.

# Introduction

The recent improvements in Artificial Intelligence (A.I.) for object recognition is largely attributable to the emergence of so-called deep learning artificial neural networks. Deep learning has developed as the natural progression from 'shallow networks' to multi-layered networks of neurons that are able to transform representations (of data, including images) from simple to complex, with increasing layer depth [1]. A review of deep learning and its broader capabilities was presented by LeCun et al. [2], and a review summarising the utilisation of machine learning in the study of materials degradation was also recently reported [3]. The automated detection of corrosion from images, either photographs or video, which may be collected from simple or sophisticated devices (i.e., smart phones, drones, etc.), presents significant advantages in terms of corrosion monitoring. Such advantages include ready access to geographically remote locations, the mitigation of safety risks to inspectors, cost savings and in monitoring speed. The automated detection of corrosion requires deep learning to approach human level intelligence. In order to demonstrate proof of concept the current research is specifically focused on the detection of ferrous corrosion, commonly referred to as 'rust'. A number of emerging, and presently evolving, technologies have enabled the development of deep learning methods, including the increased computing power of graphical processing units (GPUs), algorithm developments, and the capability to build large training sets via the Internet.

Supervised learning A.I. utilises 'labelled data' to train neural networks. Such data, which could be an image, will include identification of whether or not corrosion is present in that image. During training an objective function is optimized by adjusting weights of neurons in a neural network [4]. In the case of convolutional neural networks, the neurons are arranged into layers of filters, that convolve the input and then output a non-linear transform of the data. Fundamentally, supervised learning drives the model to develop its own filters to maximize its probability of succeeding at the task. This is an important distinction from traditional classifiers that rely on hand-coded feature detectors. In other words, A.I. image analysis to detect corrosion does not require hard coding, but a deep neural network 'learns' to identify corrosion through its own algorithms. However, one of the major drawbacks of supervised learning is the need for labelled training data. In general terms, more training data leads to better A.I. accuracy [5–8]. Furthermore, it has been demonstrated that using more training data outperforms A.I. models developed with more accurately labelled data; provided that the incidence of so-called adversarial labelling (i.e., incorrect labelling of training data) is low [9–11]. It is believed that this is because there is a need for a certain amount of data for the model to learn a good generalized relationship between the inputs and the desired outputs.

In the interest of driving research into A.I. research, certain communities or research groups are releasing will release ready-made datasets, which include labelled data. Some of the largest and most well-known datasets of labelled images (noting that they do not include labelling for corrosion) include ImageNet [12], MS COCO [13], and ADE 20K [14,15]. It is instructive to look at the effort involved in the creation of these datasets. ImageNet consists of 3.2 million images with a per picture label. These images were collected from the Internet and labelled using Amazon Mechanical Turk, where participants from across the globe vote on the label of each image, a label was not assigned until at least 10 votes were registered. Thus, ImageNet required at least 32 million online votes for labelling. From ImageNet a subset was used for the ImageNet Large Scale Visual Recognition Challenge (ILSVRC), where a monetary prize is awarded to the most accurate automated detection model [16]. The ILSVRC has proven very successful in motivating research into A.I. for image classification.

Without the benefit of a publicly available dataset, labelling large quantities of data is the first and most important step toward developing accurate deep learning models. Methods to predict the required dataset size for a desired accuracy have been developed [17,18] that help researchers forecast the effort involved. Specifically, for corrosion detection and segmentation (i.e., per pixel labelling) the present authors found that there is a need for > 65,000 labelled images required to achieve an essentially human level accuracy for an A.I. model [11]. To generate a large labelled dataset, a special purpose website was developed at www.corrosiondetector.com. This intended to increase the rate of training by enlisting the public to both provide images and to contribute to labelling pre-existing data. Not only would this allow the developed A.I. model to approach enhanced accuracies in a shorter time frame, given that the labelling task is beyond the scope of one individual, but also provides a free online tool for image assessment.

**Methodology**

The 'Corrosion Detector' website was set up in order to conduct two separate tasks:
1. Collect votes from users to label images by indicating if they contain corrosion, or no corrosion.
2. Collect uploaded images from users (which the users can also test for the presence of corrosion by the A.I. model that is being developed by point 1).

The images presented for voting initially consisted of 859 images, which were either provided by the developers, or scraped from Google Images using keyword searches that includes terms such as 'forest', 'basketball', 'car', as well as 'corrosion' and 'rust'. Users of 'Corrosion Detector' are presented with four images at a time, selected at random, and asked to check a box next to each image that they believe shows any signs of corrosion. These votes were collected, and the majority response following 5 votes was used as the image label.

As noted in point 2. above, to encourage the uploading of images by users, users may provide photos for the deep learning model to perform inference upon (i.e., to test if the uploaded image contains corrosion). The model utilised herein to perform the tasks required of 'Corrosion Detector' was programmed in python, primarily using the PyTorch API (pytorch.org). The model architecture employed a five layer convolutional network, shown diagrammatically in Figure 1 (and modelled on the so-called encoding side of the U-Net architecture [19]).

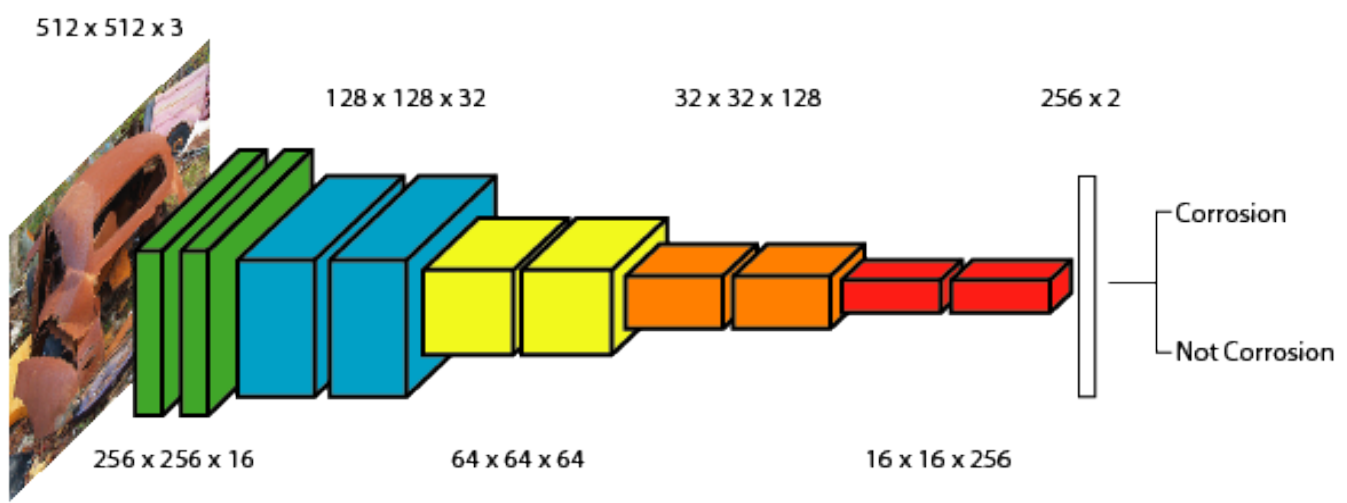

*Figure 1. Corrosion detector network architecture.*

In order to provide the reader with the ability to reproduce the corrosion detector presented in this work, the model is outlined herein, with the understanding that the details will be of most interest to a subset of the readership. Nonetheless, such details are essential to the presentation of the complete study. To undertake training, the model was initialized using Kaiming uniform initialization [20], utilizing cross-entropy loss and the ADAM optimizer with learning rate of $1 \times 10^{-4}$. Training was undertaken via the PyTorch* training scheme as follows:

```
for epoch in range(num_epochs):
    for phase in ['train', 'val']:
        if phase == 'train':
            scheduler.step()
            model.train(True)  # Set model to training mode
        else:
            model.train(False)  # Set model to evaluate mode
        # Iterate over data
        for data in dataloaders[phase]:
            inputs, labels = data
            optimizer.zero_grad()
            # forward
            outputs = model(inputs)
            _, preds = torch.max(outputs.data, 1)
            loss = torch.nn.CrossEntropyLoss()(outputs, labels)
            if phase == 'train':
                loss.backward()
                optimizer.step()
```

The baseline model (i.e., at launch of the 'Corrosion Detector' website and to serve as the baseline A.I. model for which subsequent improvements in accuracy with training could be assessed) was trained for 1 epoch, where each epoch is defined as a full pass over the dataset. Subsequently, the model was then retrained in three subsequent training sessions on the increasing dataset sizes for 25 epochs. Training was performed on Ubuntu Linux, with an NVIDIA GTX 1080Ti and CUDA acceleration.

The initial 'training set' was comprised of 600 images, which continued to increase as votes and uploaded images were added. The 'validation set' was comprised of 444 images, and the same set was used for all classification accuracy evaluation. The A.I. model accuracy was reported as the percentage of correct predictions on the entire validation set.

To run inference on user uploaded images the model was converted to ONNX format. In order for the model to operate autonomously in a cloud-based service, the entire runtime was initialized in a virtual environment with all the necessary dependencies in order to run on the commercial AWS (Amazon® Web Services) Lambda. The AWS Lambda service (which was the service selected by the authors to host 'Corrosion Detector') limited the total compressed runtime environment size to less than 50 Mb. A website 'front end' for the model was developed to provide an interface for members of the public to vote and upload images, which is the front-end known as the 'Corrosion Detector' website.

The website provides two actions for users (as depicted by Figure 2), the first is to vote on a quiz that prompts people to *'Please select the images you see **any** signs of corrosion in.'*; while the second action asks for uploaded images to *'Help Corrosion Detector train by adding images to our database!'* When a user uploads an image the most recent cloud-based evolution of the corrosion detector deep learning model reports if it detects corrosion - whereupon the user can correct erroneous detection. The uploaded images are added to the quiz, and after a minimum of five votes are eligible to be incorporated into the training set.

---

* PyTorch is a machine learning library based on the 'Torch' library, used for deep learning applications and was primarily developed by Facebook's (Facebook, Inc.) artificial intelligence research group.

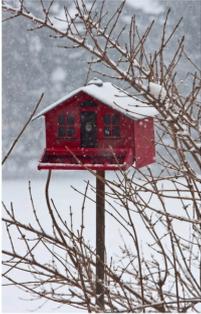

Figure 2. The two main functions of www.corrosiondetector.com; (Top) The quiz allows users to label data for training, and (Bottom) User uploaded images are assessed by the deep learning model while increasing the size of the corrosion dataset. (All images are publicly available from Google Images or provided by users with consent for use. All images at all times remain the copyright of the Author/Owner and are not to be distributed or copied).

**Results and Discussion**

*Website*

The website ('Corrosion Detector') was launched on September 26th, 2018. Its existence was and publicized globally via social media (LinkedIn, Twitter, Facebook, and via bulk email), to generate traffic and use. Over a period of 37 days 2,539 votes were recorded for labelling of images as either containing 'corrosion' or 'no corrosion'. During the same period users uploaded 310 images that were assessed online by the most recent version (at the time of upload) of the aforementioned cloud based deep learning model for the presence of corrosion. The pattern of cumulative uploads is presented in a.

The ideal target audience for the Corrosion Detector website is either corrosion researchers, corrosions specialists, or those with some capability of visually detecting corrosion (from prior experience)– with the intent of making use of their 'expert knowledge'. However, in reality, users (i.e., voters / labellers) may have included children or non-specialists. User identification was not tracked for privacy reasons, and therefore the voting and upload results are taken *prima facie*. It may however be assumed that users operate with self-selection bias, in that to sufficiently engage with the website and vote, users are likely to have some interest in corrosion - although it is not possible to validate their expert knowledge. For this reason, the commercial outsourcing of labelling (which is common in the field of machine learning for labelling of common items, such as the labelling of images with cats), was deemed inappropriate for labelling of corrosion.

*Model Accuracy*

During the period studied herein (37 days) the corrosion classification model was progressively trained based on the cumulative voting of the training set. Training was based on the 2,539 crowd-sourced votes. Relative to the validation images (which do not form part of the model training and expertly labelled for the purposes of identifying the model error), the measured classification accuracy of the model improved from 66% (day 1 of website launch) to 93% (37 days after website launch), presented in b. Following the trend of model accuracy improvement as seen in Figure 3b, to achieve a 99.9% accuracy, a total of ~30,000 votes would be required (based on the empirical evidence from the present work). This large number of required votes is, in part, the rationalisation of the crowd sourced voting sought by the Corrosion Detector website.

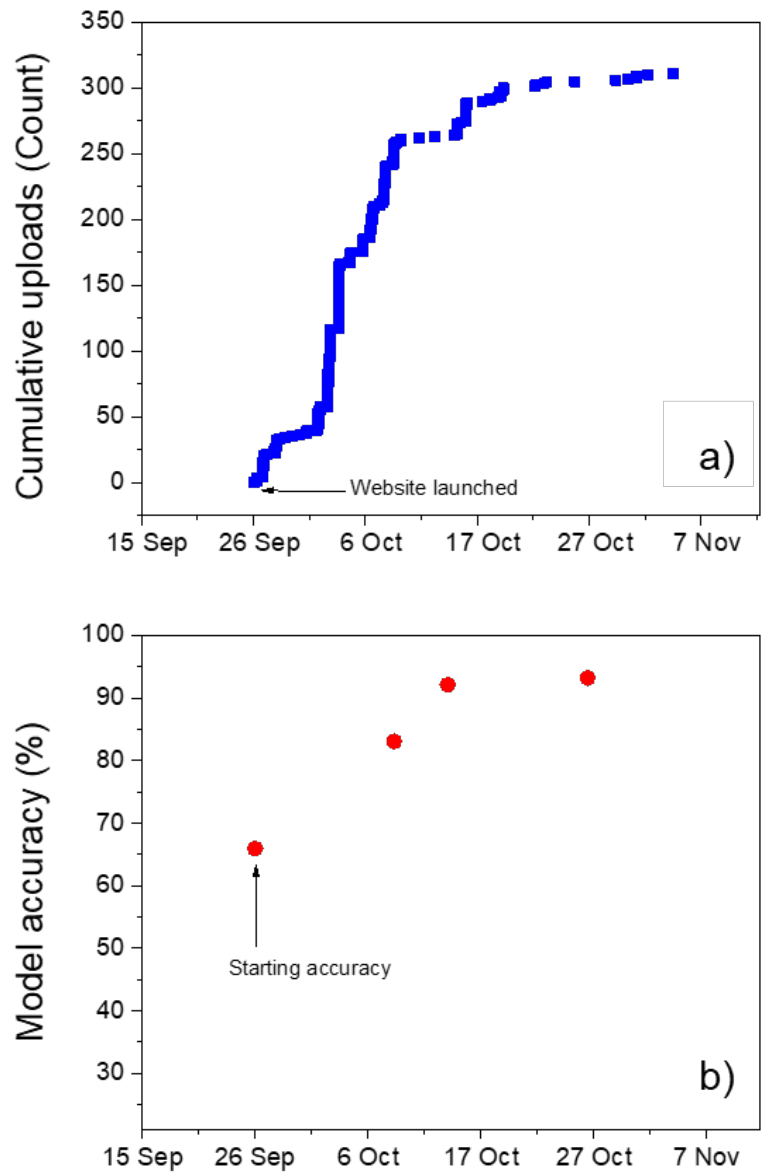

*Figure 3. (a) Cumulative uploaded contributions over time, (b) Accuracy of the model at predicting corrosion/no corrosion on the reserved validation set, reported after each periodic validation session.*

### *Voting Patterns*

Analysing the voting patterns of some specific (and interesting) results also provides insight into user interaction with the Corrosion Detector website. The voting patterns for a given image are accessible to the website authors. When probing the quality (i.e., level of 'expert') of the voters, it is seen that many images show disagreement in the voting between corrosion / no-corrosion votes, with approximately of 40 – 60 % disagreement. A subset of such images, and presentation of the respective voting results, are presented in Figure 4.

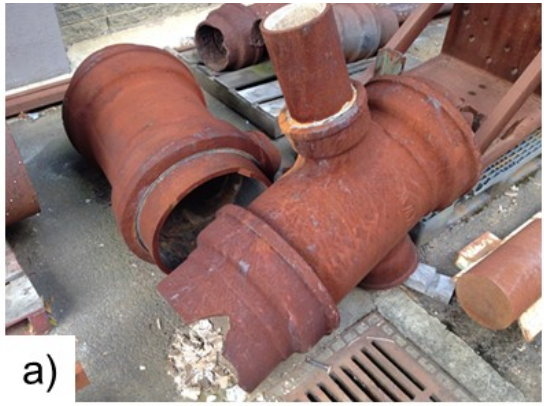
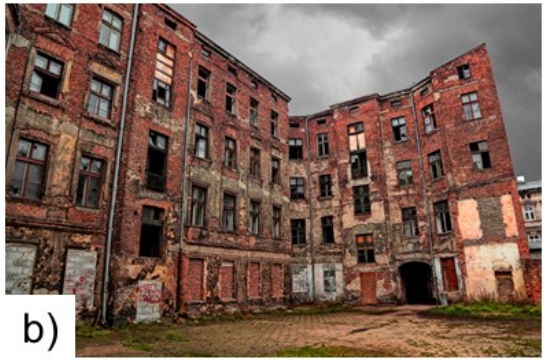
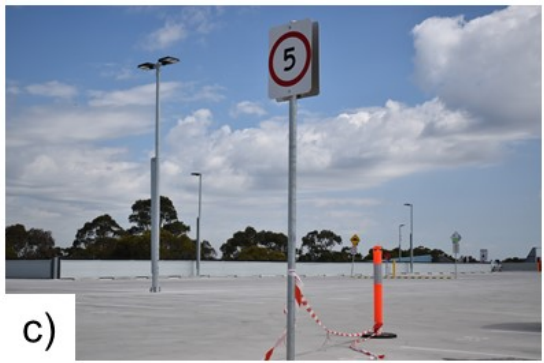
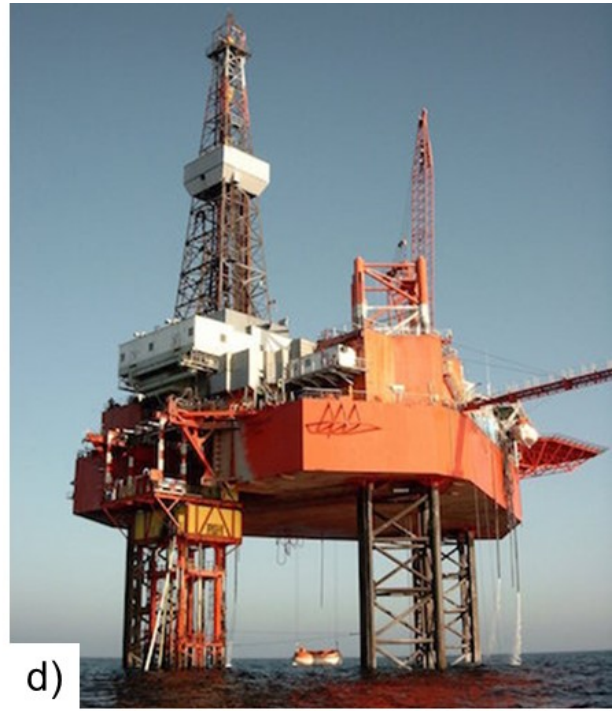
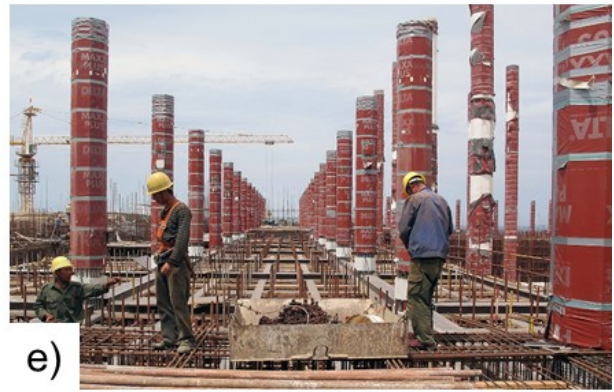

*Figure 4. (a) Image voted as containing signs of corrosion by 19 of 20 users (95%), (b) Abandoned building, this image was voted to contain signs of any corrosion by 8 out of 15 users (53%), (c) Image voted as containing no corrosion by 24 of 25 votes (4% positive for corrosion), (d) Offshore platform, this image was voted to contain any signs of corrosion by 7 of 14 users (50%), (e) Steel fixers preparing reinforcing steel, this image was voted to contain any signs of corrosion by 9 of 20 users (45%).*

The image in Figure 4a was overwhelming labelled as containing corrosion (although, one voter did not believe there was any corrosion). In Figure 4b an abandoned brick building is shown, and although there are no clearly corroded surfaces (i.e., no exposed ferrous rust), approximately 53% of voters see signs of corrosion (which suggested that users may perhaps associate deterioration or degradation of any sort, with corrosion). Figure 4c shows an empty parking lot which was overwhelming labelled as containing no corrosion - although, one voter did believe there was corrosion (suggesting that some voter(s) may neither have any experience in detecting corrosion or are not interested in faithful labelling). The offshore platform in Figure 4d shows no clear indications of corrosion, but again in this environment the presence of corrosion is inferred by 50% of the voters (even though such a critical asset is likely under a significant maintenance program and includes sacrificial anode cathode protection).

The image in Figure 4e presents steel fixers preparing reinforcing before concrete is poured. However, 9 of 20 voters decided the steel has corrosion; although those in the construction industry with experience with reinforcing steel will infer that reinforcement has mill scale, and indeed will become passive upon the imminent concrete pouring. The purpose of such elaboration for each image is to indicate that whilst images may be presented with a reasonably simple task of voting corrosion or no corrosion, human level interpretations can also include situational awareness and are uniquely based on experience. The vast majority of the community will not know that steel is passivated in contact with fresh concrete [21], nominally assuring its lack of corrosion.

The inferential presence of corrosion in the images presented in Figure 4 (and indeed, in the whole of the Corrosion Detector website) is difficult for standard deep learning models to understand – and is also frustrated by any voting discrepancies. The final example of an image and its voting (seen in ) raises a common argument regarding the nature of corrosion. Namely, the intentional iron oxide patina of weathering steel is voted as corrosion for exactly 50% of respondents. In the field of corrosion inspection, it would be 'wrong' to report this as defective, as it is an intentional feature of the product, typically selected for aesthetic purposes. However, such scenarios present themselves to corrosion inspectors daily, and therefore cannot be considered irrelevant in the development of a deep learning model. Such nuances are described further below.

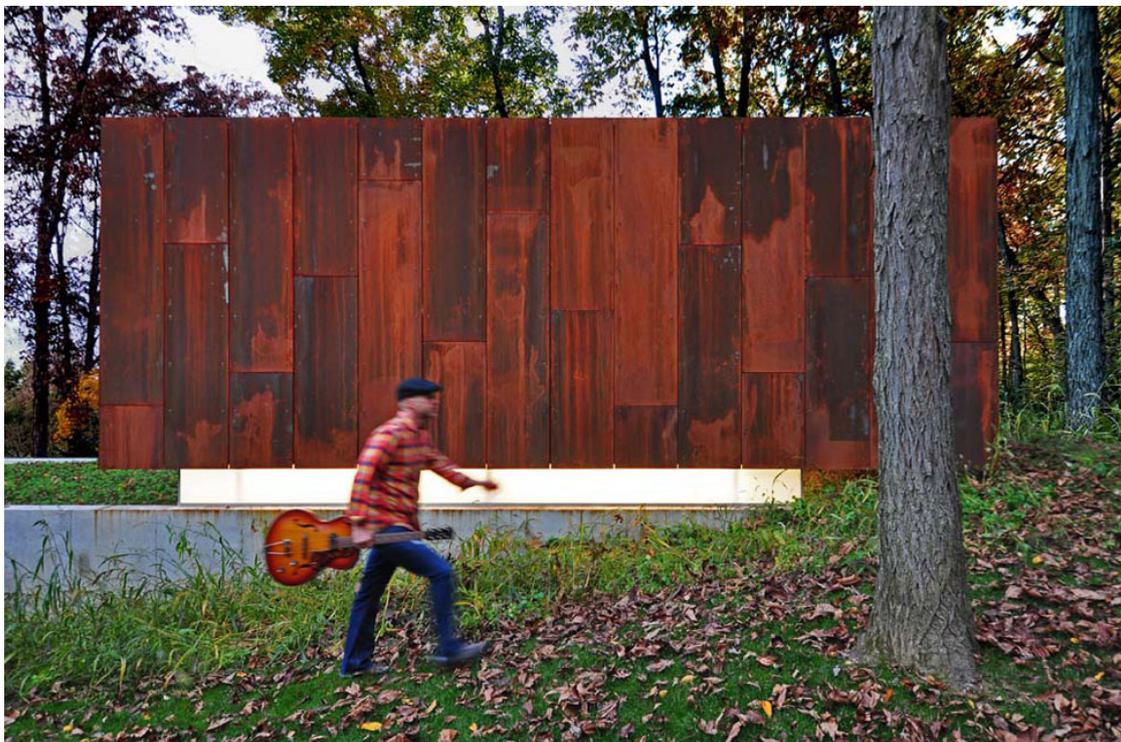

*Figure 5. Man marching with guitar in front of weathering steel, this image was voted to contain any signs of corrosion by 12 of 24 users (50%).*

### *General discussion*

The analytical results presented in Figure 3 reveal that with continued user votes (and uploads), the 'training' of a machine learning model was able to achieve a 93% accuracy over a period of 37 days. Nonetheless, it was noted that to achieve higher accuracy (such as 99.9%) that may be more typical of accuracy sought for cases where remote inspections incur significant cost, it was shown that significantly more user voting is required.

The rate of model accuracy improvement is inherently also linked to the quality of the image labelling (i.e., user voting). The ability to de-couple the effect of voting quality on model accuracy is not possible, however it is possible to probe user voting tendencies in some more detail. Probing user tendencies was done by the authors for a limited number of scenarios presented in Figures 4 and 5, and more generally for images not reported. Such tendencies, which are in part evident by the associated voting results in Figure 4, lead to the ability to discuss some salient points related to user voting.

It is evident to the authors that a portion of voters are not experts in corrosion detection, and this may include children or those particularly unskilled in visualising corrosion – and hence are unable to properly label images. This implies that there is in fact, error in the labelled dataset that is used for training. This error is an inherent aspect of crowd-sourced labelling.

It is also evident from voting tendencies, that the identification of corrosion is a highly complex task. It requires experts, and experts make numerous subconscious decisions before making an assessment. This implies that expert knowledge and decisions are based on knowledge that is *not* in the images presented in in many cases, and may include:

- Knowledge of the range of materials in an image. Expert labellers can distinguish wood from metal, and also familiar with ceramics and organic material. Expert labellers know that only metals can undergo corrosion (in the context sought), and have an exclusion filter built in based on subconscious knowledge of material types.
- Knowledge of the criticality of a situation. A superficial blemish on a non-structural component is negligible, as is surface blemishes on components still in the construction phase (such as Figure 4e), as such components will passivate post-construction completion.
- The ability to assess the difference between what may be brown or red paint, and corrosion (as corrosion will typically not appear in scenarios where there is no mechanistic reason for it to prevail).
- Situational context, such as that in image in Figure 5, where there is deliberate or inconsequential corrosion (and noting that it was voted as no-corrosion by 50% of voters)

The above aspects indicate just how complicated the labelling of corrosion can be, and how valuable expert corrosion labelling (and the inherent knowledge it entails), is. In fairness to participants of all experience levels, the quiz questions (i.e., images for user labelling, which may also be entirely random or complex images uploaded by users) may present a range of challenges in their interpretation. As a consequence, ongoing research may benefit from rephrasing the quiz query as: '*please select images that indicate steel corrosion that you would recommend be repaired*', and potentially include a self-reported level of corrosion expertise. However, the overall conclusion that corrosion can mean different things to different people remains, and the honesty of the quiz participant is an uncontrolled variable.

In the context of model accuracy, the model training is also influenced by the images being labelled (including the no corrosion images). The ability to have an ideal dataset would mimic the images that humans experience in their daily lives (such as what we see from when we wake to when we go to sleep). The images associated with Corrosion Detector are at present, skewed on the basis that there are many images with corrosion, relative to no corrosion – however the latter which is far more common in reality. In our daily activities, the number of images we see are not typical of corrosion contacting. Also, the extent of corrosion we encounter will be significantly impacted by where we live (what country) and in what industry we work. As such, the generation of a balanced dataset – publicly available -also remains an issue that should be addressed by the broader technical community.

Conclusions

- The work presented herein reveals the successful establishment of the Corrosion Detector Website, created for users to vote on (i.e., label) whether images contain corrosion or no corrosion. This voting was successfully translated to a machine learning model to produce an automated corrosion detection model.
- Using crowd-sourced voting, 2,539 votes allowed the machine learning model to improve its accuracy of corrosion detection from 66% to 93% in 37 days.
- Rationalisation of the performance of the machine learning model presented was probed by exploring the voting preference for specific images. Certain images proved controversial, with 40 – 60% of voters disagreeing. Typically, such images may not explicitly show corrosion, but rather the context suggests that corrosion may be present.
- A discussion was elaborated, highlighting the complexity of corrosion detection from images, and where expert corrosion labellers may deviate (at times significantly) from non-expert labellers, based on their situational and materials awareness. Nonetheless, the presented machine learning model, and its ability to automatically detect corrosion from user images, presents a significant prospect in the context of future maintenance and asset management.

**Acknowledgements**
We acknowledge support from Woodside Energy.